\documentclass[journal,draftcls,onecolumn,12pt]{IEEEtran}
\usepackage{mathrsfs}
\usepackage{pifont}
\usepackage{bbding}
\usepackage{amsmath,epsfig}
\usepackage{stmaryrd}
\usepackage{amssymb}
\usepackage{amsfonts}
\usepackage{epic}
\usepackage{graphicx}
\usepackage{subfigure}
\usepackage{curves}
\usepackage{enumerate}
\usepackage{algorithm}
\usepackage{algpseudocode}

\usepackage{stfloats}
\usepackage{latexsym}
\usepackage{epstopdf}
\usepackage{epic}
\usepackage{multirow}
\usepackage{stfloats}
\usepackage{bm}

\usepackage{booktabs}

\begin{document}

\newcommand{\reffig}[1]{Fig. \ref{#1}}

\newtheorem{definition}{Definition}
\newtheorem{theorem}{Theorem}
\newtheorem{corollary}{Corollary}
\newtheorem{lemma}{Lemma}

\renewcommand{\algorithmicrequire}{\textbf{Input:}}  
\renewcommand{\algorithmicensure}{\textbf{Output:}}  
\newenvironment{sequation}{\begin{equation}}{\end{equation}}
\newcounter{TempEqCnt}

\title{Compressed Channel Estimation with Position-Based ICI Elimination for High-Mobility SIMO-OFDM Systems}

\author{Xiang Ren,  Meixia Tao, \emph{Senior Member, IEEE},\\and Wen Chen, \emph{Senior Member, IEEE}

\thanks{The authors are with Department of Electronic Engineering,
Shanghai Jiao Tong University, China (e-mail: \{renx, mxtao, wenchen\}@sjtu.edu.cn).
}}

\maketitle

\begin{abstract}
Orthogonal frequency-division multiplexing (OFDM) is widely adopted for providing reliable and high data rate communication in high-speed train systems.
However, with the increasing train mobility, the resulting large Doppler shift introduces intercarrier interference (ICI) in OFDM systems and greatly degrades the channel estimation accuracy.
Therefore, it is necessary and important to investigate reliable channel estimation and ICI mitigation methods in high-mobility environments.
In this paper, we consider a typical HST communication system and show that the ICI caused by the large Doppler shift  can be mitigated by exploiting the train position information as well as the sparsity of the conventional basis expansion model (BEM) based channel model.
Then, we show that for the complex-exponential BEM (CE-BEM) based channel model, the ICI can be completely eliminated to get the ICI-free pilots at each receive antenna.
After that, we propose a new pilot pattern design algorithm to reduce the system coherence and hence can improve the compressed sensing (CS) based channel estimation accuracy.
The proposed optimal pilot pattern is independent of the number of receive antennas, the Doppler shifts, the train position, or the train speed.
Simulation results confirms the performance merits of the proposed scheme in high-mobility environments.
In addition, it is also shown that the proposed scheme is robust to the respect of high mobility.

\end{abstract}


\section{Introduction}
High speed trains (HST)  have been increasingly developed in many countries and especially have made great impact in China.
There is a growing demand of offering passengers the data-rich wireless communications with high data rate and high reliability.
Orthogonal frequency-division multiplexing (OFDM), as a leading technique in the current Long Term Evolution (LTE) and  future evolution of cellular networks, have demonstrated great promise in achieving high data rate in stationary and low-mobility environment.
In the HST environment, however, since the train travels at a speed more than 350km/h, the high Doppler shift  destroys the orthogonality resulting in the intercarrier interference (ICI) in OFDM systems.
This directly degrades the channel estimation accuracy and significantly affects the overall system performance.
It is thus necessary and important to investigate reliable channel estimation and ICI mitigation methods in high-mobility environments.

Channel estimation in OFDM systems over time-varying channels has been a long-standing issue \cite{3}-\cite{14}.
The existing works can be generally divided into three categories based on the channel model properties they adopted.
The first category of estimation methods adopted the linear time-varying channel model, i.e., the channel varies with time linearly in one or more OFDM symbols, such as \cite{3} and \cite{4}.
The second category employs the basis expansion model (BEM) such as \cite{5}-\cite{7}.
Note that both these two channel models implicitly assume that the channel is in rich-scattering evironment with sufficient multipath.
The third category of channel estimation methods is based on the recent research finding that
wireless channels tend to exhibit sparsity, where the channel properties are dominated by a relatively small number of dominant channel coefficients.
Thus, to utilize the channel sparsity, several works \cite{8}-\cite{14}  studied the applications of compressed sensing (CS) in the channel estimation over  doubly-selective channels.

Another line of research to improve the performance of channel estimation is to consider the pilot design and channel estimation jointly.
Recently, many researches considered this problem based on the CS-based channel estimation methods.
Coherence is a critical metric in CS as it directly influences the CS recovery performance  \cite{15}-\cite{17}.
The works \cite{15} and \cite{16} concluded that a lower system coherence leads to a better recover performance.
Based on these results, previous works \cite{18}-\cite{26} proposed several pilot design methods to reduce the system coherence and hence to improve the CS-based channel estimation performance.
The works \cite{18}-\cite{20} proposed the pilot pattern design methods based on the exhaustive search to reduce the CS coherence for single-input single-output (SISO) and multiple-input multiple-output (MIMO) systems. These methods, however, need large iteration times to achieve satisfactory estimation performance. The works \cite{21} and \cite{22} utilized the discrete stochastic approximation to design the pilot pattern for OFDM systems. These  above works designed the pilot pattern and assumed that the pilot symbols are the same.
In our previous work \cite{23}, we proposed a pilot symbol design method with the equidistant pilot pattern for high-mobility MIMO-OFDM systems.
In our previous work \cite{25}, we proposed a position-based joint pilot pattern and pilot symbol design method for the high-mobility OFDM systems, where different pilots are designed for the Doppler shifts at different train positions and then stored into a codebook.
For each train position, the system selects the corresponding optimal pilot and uses it to estimate the channel. However, none of the above mentioned works considered the ICI mitigation. In the presence of a high Doppler shift, the ICI can reduce the channel estimation performance significantly.
In specific, the work \cite{26} proposed a pilot pattern design method with the ICI-free structure for the distributed compressed sensing (DCS) based channel estimator over doubly-selective channels. However, this method needs a large number of guard pilots to eliminate the ICI, which reduces the spectral efficiency.

This work is based upon our previous work \cite{25} with the aim of improving the channel estimation performance by taking ICI into account. Similar to \cite{25}, we consider a HST communication system where the instantaneous position and speed of the moving train can be estimated, for example using a global positioning system (GPS). But different from \cite{25}, we consider the single-input multi-output (SIMO) scenario.
Note that the method in \cite{25} cannot be directly applied to SIMO systems.
The optimal pilot proposed in \cite{25}  for the SISO system is different for different train positions.
When it applied in the SIMO system, different optimal pilots need to be sent for different receive antennas due to their different positions. This will certainly reduce the spectral efficiency.

In this paper, based on the conventional BEM, we first show that the ICI caused by the large Doppler shift  can be mitigated by exploiting the train position information. The relationships between the dominant channel model, the dominant channel coefficients, the Doppler shift, and the train position are also given. Then, considering the complex-exponential BEM (CE-BEM), we propose a new low complexity position-based ICI elimination method, by which we can get the ICI-free pilots at each receive antenna.
In contrast to the conventional iterative ICI mitigation method in \cite{32}, the proposed method only requires a permutation of the received subcarriers, which is much less complex. In addition, different from the methods in \cite{7} and \cite{26} needing large number of guard pilots to eliminate the ICI, the proposed method does not need any guard pilot, which highly improves the spectral efficiency as well.
After getting the ICI-free pilots,  we formulate the pilot pattern design problem to minimize the system average coherence, and propose a new pilot design algorithm to solve it.
Specifically, the optimal pilot pattern is independent of the train speed, the train position, the Doppler shift, or the number of receive antennas.
Thus, the system only needs to store one pilot pattern, which highly reduces the system complexity in contrast to \cite{25} which selecting different pilots for different train positions.
In addition, different from the methods in \cite{25} and \cite{32} that the channel estimation performances are highly influenced by the system mobility, simulation results demonstrate that the proposed scheme is robust to the high mobility.

The rest of this paper is organized as follows. Section II introduces the HST communication system, the SIMO-OFDM system model, and  the conventional BEM. In Section III,  we exploit the position information of the BEM. Then, for the CE-BEM, we introduce a new position-based ICI elimination method. In Section IV,  after briefly review some CS fundamentals, we formulate the pilot design problem and propose a new low coherence pilot pattern design algorithm.  The complexity and the practical applicability of the proposed scheme is also discussed.  Section V presents simulation results in the high-mobility environment. Finally, Section VI concludes this paper.

$Notations$:
$\left\|\cdot\right\| _{\ell _0 }$ denotes the number of nonzero entries in a matrix or vector, and $\left\|\cdot\right\| _{\ell _2 }$ is the Euclidean norm.
${\bf X} ({\bf w},:)$ denotes the rows of the matrix ${\bf{X}}$ whose row indices are in the vector $\bf w$.
The superscripts $(\cdot)^T$ and $(\cdot)^H$ denote the transposition and Hermitian of a matrix, respectively. $\lceil\cdot\rceil$ denotes the round up operator,  $\lfloor\cdot\rfloor$ denotes the round down operator,
$\otimes$ denotes the Kronecker product,  and $\text{diag}\{\cdot\}$ denotes the operator that changes a vector to a diagonal matrix.
${\bf I}_K$ denotes the $K\times K$ identity matrix, and ${\bf I}^{\langle q\rangle}_K$ denotes a permutation matrix which is obtained form ${\bf I}_K$ by shifting its column circularly $|q|$ times to the right for $q<0$ and to the left otherwise.
Finally, $\mathbb{C}^{M\times N}$ denotes the set of $M\times N$ matrices in the complex field, and $\mathbb{R}$ denotes the real field.

\section{System Model}

\subsection{HST communication system}
We consider a typical broadband wireless communication system for high speed trains (HST) \cite{25}\cite{27}, as shown in Fig.~1.  The communication between the base stations (BS) and the mobile users is conducted in a two-hop manner through a relay station (RS) deployed on the train.
The RS is connected with several antennas evenly located on the top of the train to communicate with the BS.
Moreover, the RS is also connected with multiple indoor antennas distributed in the train carriages to communicate with mobile users by existing wireless communication technologies. The BSs are located along the railway at some intervals and connected with optical fibers. Here we assume that each BS is equipped with one antenna for simplicity and has the same transmit power and coverage range.
Similar to \cite{25}, we assume that the HST is equipped with a GPS which can estimate  the HST's instant position and speed information perfectly and send them to the RS with no time delay \cite{28}. But, different from \cite{25}, in this paper, the BS does not need to receive these information from the GPS.

\begin{figure}[t!]
  \centering
  \includegraphics[width= 5in]{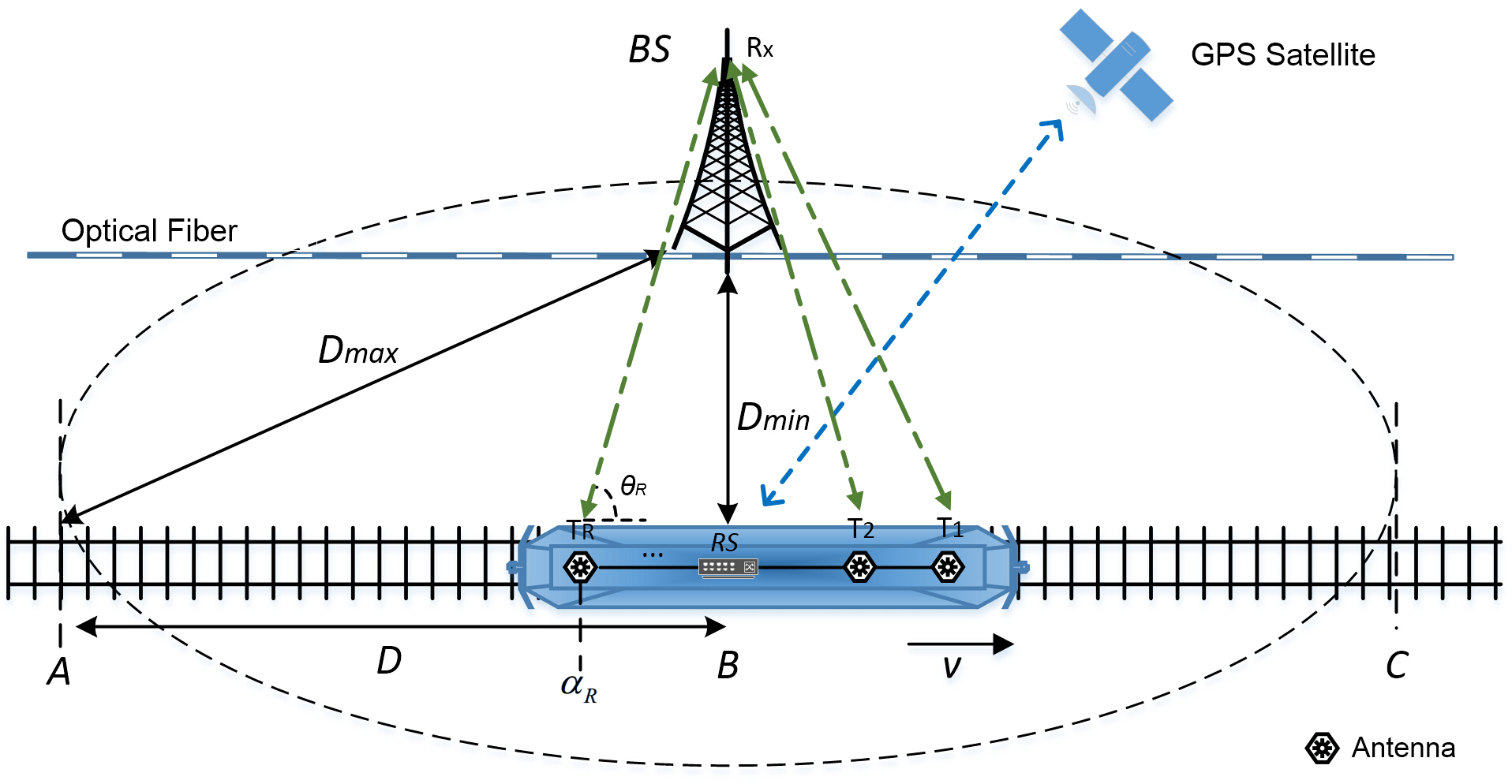}\\
  \caption{The structure of a multi-antenna HST communication system.}\label{sys}
\end{figure}

In \reffig{sys}, we assume that the HST is traveling towards a fixed direction at a constant speed $v$.
Let $D_{max}$ denote the maximum distance from the BS to the railway (i.e., the position $A$ and $C$ to $BS$), $D_{min}$ denote the minimum distance (i.e., the position $B$ to $BS$), and $D$ denote the distance between $A$ and $B$.
The $R$ receive antennas evenly located on the top of the HST are denoted as $\{T_r\}^R_{r=1}$,
and $R_x$ denotes the antenna equipped on the BS.
In each cell, we define $\alpha_r$ as the distance between the $r$-th receive antenna and the position $A$, and define $\theta_r$ as the angle between the BS to $T_r$ and the railway.
When the HST moves from $A$ to $C$,  $\theta_r$  changes  from $\theta_{min}$ to $\theta_{max}$.
If $D_{max} \gg D_{min}$, we have $\theta_{min}\approx 0^\circ$ and $\theta_{max}\approx 180^\circ$.
For $T_r$ at a certain position $\alpha_r$, it suffers from a Doppler shift $f_{r}$, where $f_{r}$  can be calculated by $f_{r} = \frac{v}{c}\cdot f_c \cos\theta_r$ with the carrier frequency $f_c$  and the the light speed $c$.
In addition, we assume that $f_{r}$ is constant within one OFDM symbol.

\subsection{SIMO-OFDM system} \label{SIMO}


In this paper, we only consider the first-hop communication in the HST system, i.e., the communication from the BS to the RS. It is treated as  a SIMO-OFDM system with one transmit antenna and $R$ receive antennas. Suppose there are $K$ subcarriers.
The transmit signal at the $k$-th subcarrier during the $n$-th OFDM symbol is denoted as $X^n (k)$, for $n = 1,2,...,N$ and $k = 1,2,...,K$. At the BS, after passing the inverse discrete Fourier transform (IDFT) and inserting the cyclic prefix (CP),  the signals are transmitted to the wireless channel.
At the $r$-th receive antenna, after removing the CP and passing the discrete Fourier transform (DFT) operator, the received signals  in the frequency domain are represented as
\begin{equation}
\mathbf{y}^n_r  =\mathbf{H}^n_r\mathbf{x}^n+\mathbf{n}^n_r, \label{eq01}
\end{equation}
where ${\bf{y}}^n_r = [Y^n_r(1),Y^n_r(2),...,Y^n_r(K)]^T$ is the received signal vector over all subcarriers during the $n$-th OFDM symbol, ${\bf{H}}^n_r$ is the channel matrix between the transmit antenna and the $r$-th receive antenna,
${\bf{x}}^n= [X^n(1),X^n(2),...,X^n(K)]^T$ is the transmitted signal vector,
and ${\bf{n}}^n_r = [N^n_r(1),N^n_r(2),...,N^n_r(K)]^T$ denotes the noise vector, where $N^n_r(k)$ is the additive white Gaussian noise (AWGN) with a zero mean and $\sigma^2_\varepsilon$ variance.

If the channel is time-invariant,
the off-diagonal term ${{H}_r^n({k,d}})$ $(k\neq d)$ in ${\bf{H}}_r^n$ is negligible, and the diagonal term ${{H}^n_r({k,d}})$ $(k=d)$ alone represents the channel, where $k,d = 1,...,K$.
However, for time-varying channel, the off-diagonal term cannot be neglected and (\ref{eq01}) can be rewritten as
\begin{align}
{\bf{y}}^n_r  &={\bf{H}}_{r_{\rm free}}^n{\bf{x}}^n + {\bf{H}}_{r_{\rm{ICI}}}^n{\bf{x}}^n + {\bf{n}}_r^n,\label{eq02}
\end{align}
where ${\bf{H}}_{r_{\rm free}}^n \triangleq {\text{diag}}\{[H^n_r(1,1),H^n_r(2,2),...,H^n_r(K,K)]\}$ denotes the ICI-free channel matrix, and ${\bf{H}}_{r_{\rm ICI}}^n \triangleq {\bf{H}}^n_r - {\bf{H}}_{r_{\rm free}}^n$ is the ICI part.

%

\subsection{BEM-based Channel Model}
In our previous work \cite{25}, we adopted the channel model in \cite{10} to model the channel in the delay-Doppler domain.
In this work, however, we employ the BEM to model the high-mobility channel in the time domain.
Assume that the channel between the transmit antenna and each receive antenna consists of $L$ multi-paths.
For each channel tap $l$, $0\leq l\leq L-1$, we define ${\tilde{\bf{h}}}^n_r(l)= [h^n_r(0,l), h^n_r(1,l), ..., h^n_r(K-1,l)]^T\in\mathbb{C}^{K\times1}$ as a vector which collects the time variation of the channel tap within the $n$-th OFDM symbol of the channel between the transmit antenna and the $r$-th receive antenna. Denote $f_{max}$ as the maximum Doppler shift, $T$ as the packet duration, and $Q = 2\lceil f_{{max}} T\rceil$ as the maximum number of the BEM order.
Then, each $\tilde{\bf{h}}^n_r(l)$ can be represented as
\begin{align}
  \tilde{{\bf{h}}}^n_r(l)= {\bf B}{\bf c}^n_r(l) + {\bm{\epsilon}}^n_r(l),
\end{align}
where  ${\bf B} = [{\bf b}_{0},...,{\bf b}_q,...,{\bf b}_{Q}]\in\mathbb{C}^{K\times (Q+1)}$  collects $Q+1$ basis functions as columns,  ${\bf b}_q$ denotes the $q$-th basis function ($q = 0,1,...,Q$) whose expression is related to a specific BEM model,  ${\bf c}^n_r(l) = [c^n_r(0,l),c^n_r(1,l),...,c^n_r(Q,l)]^T$ represents the BEM coefficients for the $l$-th tap of the channel at the $r$-th receive antenna within the $n$-th OFDM symbol, and ${\bm \epsilon}^n_r(l) = [\epsilon^n_r(0,l),\epsilon^n_r(1,l),...,\epsilon^n_r(K-1,l)]^T$ represents the BEM modeling error.

Stacking all the channel taps of the $r$-th receive antenna within the $n$-th OFDM symbol in one vector
\begin{equation}
\tilde{{\bf h}}^n_r = [{h}^n_r(0,0),...,{h}^n_r(0,L-1),...,{h}^n_r(K-1,0),...,{h}^n_r(K-1,L-1)]^T \in \mathbb{C}^{KL \times 1},
\end{equation}
then we  obtain
\begin{equation}
  \tilde{{\bf h}}^n_r = ({\bf B} \otimes {\bf I}_L){\bf c}^n_r+{\bm \epsilon}^n_r, \label{eq03}
\end{equation}
where ${\bf I}_L$ is an $L\times L$ identity matrix, ${\bf c}^n_r = [c^n_r(0,0),...,c^n_r(0,L-1),...,c^n_r(Q,0),...,c^n_r(Q,L-1)]^T\in\mathbb{C}^{L(Q+1)\times 1}$ is the  stacking coefficient vector, and ${\bm \epsilon}^n_r= [\epsilon^n_r(0,0),...,\epsilon^n_r(0,L-1),...,\epsilon^n_r(K-1,0),...,\epsilon^n_r(K-1,L-1)]^T\in\mathbb{C}^{KL\times 1}$. In the following, as our focus is to discuss the performance of the channel estimator and the ICI eliminator, we ignore ${\bm \epsilon}^n_r$ for convenience.
We also assume that the coefficients are constant within one OFDM symbol.

Therefore, based on the BEM,  we can describe the system in the high-mobility environment. In addition, since we only consider the system in a single OFDM symbol in this paper, the symbol index $n$ is omitted in the sequel for compactness. Then, substituting (\ref{eq03}) into (\ref{eq01}), we obtain
\begin{align}
  {\bf y}_r =  \sum^Q_{q=0}{\bf D}_q{\bf \Delta}_{r,q}{\bf x} + {\bf n}_r,\label{eq10}
\end{align}
in which ${\bf D}_q = {\bf F}{\text{diag}}\{{\bf b}_q\}{\bf F}^H$ denotes the $q$-th BEM basis function in the frequency domain, ${\bf F}$ is the $K \times K$  DFT matrix,  ${\bf \Delta}_{r,q} = {\text{diag}}\{{\bf F}_L{\bf c}_{r,q}\}$ is a diagonal matrix whose diagonal entries are the frequency responses of ${\bf c}_{r,q}$,
$ {\bf c}_{r,q} = [c_r(q,0),...,c_r(q,L-1)]^T$ denotes the BEM coefficients of all taps of the $r$-th receive antenna corresponding to the $q$-th basis function, and ${\bf F}_L$ denotes the first $L$ columns of $\sqrt{K}{\bf F}$.


\section{Position-based ICI Elimination}

In this section,  we show that the ICI caused by the large Doppler shift in the CE-BEM channel model can be eliminated by exploiting the train position information. This is a key finding of this work, based on which we then propose a new pilot pattern design algorithm in the next section.

\subsection{Exploiting the position information of BEM}

We first give a definition of  $S$-sparse channels  based on the BEM channel model introduced in the previous section.

\begin{definition}[$S$-sparse Channels \cite{9}]\label{df1}
For a BEM-based channel model given in (\ref{eq03}), its dominant coefficients are defined as the BEM coefficients which contribute significant powers, i.e., $|{c_r(q,l)}|^2 > \gamma$, where $\gamma$ is a pre-fixed threshold.
We say that the channel is  $S$-sparse if  the number of its dominant coefficients satisfies $S = \left\| {{\bf{c}}}_r \right\|_{\ell _0 }  \ll N_0 = L(Q+1)$, where $N_0$ is the total number of the BEM channel coefficients.
\end{definition}

Then we give the following theorem, which  reflects the position information of the given system.

\begin{theorem}[Position-based S-sparse channels]
For the considered HST system, the high-mobility channel between the transmit antenna and each receive antenna at any given train position  is $S$-sparse.
\begin{proof}
When the HST as shown in \reffig{sys} moves to a certain position at a constant speed $v$, the $R$ receive antennas are at positions $\{\alpha_r\}^R_{r=1}$ and suffer from different Doppler shifts $\{f_{r}\}^R_{r=1}$, respectively, where each $f_{r}$ can be calculated with the known $\alpha_r$ supported by the GPS.
As assumed in the previous section, each $f_{r}$ is constant within one OFDM symbol.
Then, since the channel coefficient $c_r(q,l)$ is only related to the basis index $q$ ($q$ also represents the level of the Doppler shift $f_r$ at the $r$-th receive antenna) and the multipath index $l$, we can find that the dominant coefficients of the $r$-th receive antenna at $\alpha_r$ only exist in its  dominant subvector
 \begin{equation}
{\bf{c}}^*_{r} = {\bf{c}}_{r,q|q=q^*_r}=
\left[\begin{matrix}
   c_r(q^*_r,0), c_r(q^*_r,1), ..., c_r(q^*_r,L-1)\\
\end{matrix}\right]^T,
\end{equation}
where $q^*_r$  is called as the dominant index of the $r$-th receive antenna, and $q^*_r$ denotes the level of the Doppler shift  $f_r$ when the $r$-th receive antenna moves to the position  $\alpha_r$.
This is reasonable because, when the $r$-th receive antenna moves to $\alpha_r$, all channel taps suffer from the same Doppler shift $f_r$ and thus correspond to the same index $q = q^*_r$.  We will give the relationships between $f_r$, $\alpha_r$ and $q^*_r$ in the following part.
Moveover, as ${\bf{c}}^*_{r}$ contains at most $L$ dominant coefficients and  the sparsity is $S$, we have $\left\| {{\bf{c}}}^*_{r} \right\|_{\ell _0 } =\left\| {{\bf{c}}}_r \right\|_{\ell _0 } = S\leq L < L(Q+1)$. In addition, high-mobility channels are considered as the doubly-selective channels and have the multipath sparsity  \cite{8}-\cite{10}, which means that there are only $S$ paths $(S\ll L)$ with large coefficients while others can be neglected. Furthermore, as $Q$ increases with the high Doppler shift caused by the fast HST speed, high-mobility will introduce a large $Q$. Therefore, the high-mobility channel is $S$-sparse and we have $\left\| {{\bf{c}}}^*_{r} \right\|_{\ell _0 }  = \left\| {{\bf{c}}}_{r} \right\|_{\ell _0 } =S\ll L \ll L(Q+1)$.
\end{proof}
\end{theorem}

Accordingly, the relationship between the dominant index $q^*_r$ and $f_{r}$ is given as
\begin{align}
q^*_r=\left\{\begin{matrix}
  &\left\lceil{T}{f_{r}}\right\rceil+\frac{Q}{2},~&f_r\in\left[0,f_{{max}}\right],\\
  &\left\lfloor{T}{f_{r}}\right\rfloor+\frac{Q}{2},~&f_r\in\left[-f_{{max}},0\right).
\end{matrix}\right.\label{q_fd}
\end{align}
Denote ${F} =T  f_{{\max }} =T\frac{v}{c} \cdot f_c$. Then the relationship between  $q^*_r$ and $\alpha_r$ can be represented as
\begin{align}
q^*_r=\left\{\begin{matrix}
  &\left\lceil{F}\cdot\frac{D-\alpha_r}{\sqrt{{(D-\alpha_r)}^2+{D_{min}}^2}}\right\rceil+\frac{Q}{2},~&{\alpha_r}\in[0,D],\\ \\
   &\left\lfloor{F}\cdot\frac{D-\alpha_r}{\sqrt{{(D-\alpha_r)}^2+{D_{min}}^2}}\right\rfloor+\frac{Q}{2},~&{\alpha_r}\in(D,2D],
\end{matrix}\right.\label{q_a}
\end{align}
where ${\alpha_r}\in[0,D]$ denotes the $r$-the receive antenna moving from $A$ to $B$, and ${\alpha_r}\in(D,2D]$ denotes moving from $B$ to $C$.

From Theorem 1, we readily have the following corollary.


\begin{corollary}
For the considered HST system with any given train position, the high-mobility channel between the transmit antenna and the $r$-th ($r=1,2,...,R$) receive antenna is $S$-sparse, and
it can be modeled with its dominant coefficients and the dominant basis function, i.e.,  ${\bf H}_r  = {\bf D}^*_{r}{\bf \Delta}^*_{r}$, where ${\bf \Delta}^*_{r} = {\text{diag}}\{{\bf F}_L{\bf c}^*_{r}\}$, and ${\bf D}^*_{r} = {\bf D}_{q|q={q}^*_r} $ is  the dominant basis function of the $r$-th receive antenna. The relationships between the  dominant index $q^*_r$, the Doppler shift  $f_{r}$, and the antenna position $\alpha_r$ are given as (\ref{q_fd}) and (\ref{q_a}), respectively.
\end{corollary}

According to Corollary 1,  (\ref{eq10}) can be simplified as
\begin{align}
  {\bf y}_r  
                    & =  {\bf D}^*_{r}{\bf \Delta}^*_{r}{\bf x} + {\bf n}_r.
\end{align}
In this way, we exploit the position information of the BEM and utilize it to simply the required channel coefficients from $L(Q+1)$ to $L$.
Note that these analyses and conclusions are not restricted to any specific BEM.

\subsection{Position-based ICI Elimination}

In this subsection, we consider the CE-BEM \cite{251} due to its independence of the channel statistics and it is strictly banded in the frequency domain. Specifically, for the CE-BEM,  the $q$-th  basis function ${\bf b}_q$ can be represented as
\begin{equation}
  {\bf b}_q = {\left[ {\begin{array}{*{20}{c}}
1,& \cdots,&{{e^{j\frac{{2\pi }}{K}k(q - \frac{Q}{2})}}},& \cdots, &{{e^{j\frac{{2\pi }}{K}(K - 1)(q - \frac{Q}{2})}}}
\end{array}} \right]^T}.
\end{equation}
Then, the ${\bf D}_q$ can be written as
\begin{align}
  {\bf D}_q &= {\bf F}{\text{diag}}\{{\bf b}_q\}{\bf F}^H = {\bf I}^{\langle q- \frac{Q}{2}\rangle}_K{\bf F}{\bf F}^H,\\
                    &= {\bf I}^{\langle q- \frac{Q}{2}\rangle}_K,  \label{eq22}
\end{align}
where $ {\bf I}^{\langle q- \frac{Q}{2}\rangle}_K$ denotes a  matrix obtained from a $K\times K$ identity matrix ${\bf I}_K$ with a permutation $q-Q/2$.
Then, we have
\begin{align}
  {\bf H}_r & = \sum^Q_{q=0}{\bf I}^{\langle q- \frac{Q}{2}\rangle}_K{\bf \Delta}_{r,q}.
\end{align}
By detecting the matrix structure, we find that ${\bf H}_r$ is strictly banded with the bandwidth $Q+1$, which means that the $Q$ neighboring subcarriers give rise to interference, i.e., the desired signal suffers from the ICI from the $Q$ neighboring subcarriers.

Assume that $P~(P<K)$ pilots are inserted in the frequency domain at the BS with the pilot pattern ${\bf w}$,  where ${\bf w} =[w_1,w_2,...,w_P]$.
Denote ${\bf d}$ as the subcarrier pattern of the transmitted data.
Then, the received pilots at the $r$-th receive antenna is represented as
\begin{align}
  {\bf y}_r({\bf w}) =    \sum^Q_{q=0}{\bf D}_q({\bf w},{\bf w}){\bf \Delta}_{r,q}({\bf w},{\bf w}){\bf x}({\bf w}) + \underbrace{ \sum^Q_{q=0}{\bf D}_q({\bf w},{\bf d}){\bf \Delta}_{r,q}({\bf d},{\bf d}){\bf x}({\bf d})}_{\bf G}+{\bf n}_r({\bf w}), \label{eq26}
\end{align}
in which $ {\bf D}_q({\bf w},{\bf w})$ and ${\bf \Delta}_{r,q} ({\bf w},{\bf w})$ represent the submatrices of ${\bf D}_q$ and ${\bf \Delta}_{r,q}$ with the row indices ${\bf w}$ and the column indices ${\bf w}$, respectively, ${\bf D}_q({\bf w},{\bf d})$ and ${\bf \Delta}_{r,q} ({\bf d},{\bf d})$ represent the submatrices with the  row indices ${\bf w}$ and $\bf d$ and the column indices ${\bf d}$, respectively, and ${\bf n}_r$ is the noise vector at ${\bf w}$.
In (\ref{eq26}), we decouple the ICI caused by the $Q$ neighboring data from the pilots and put it in the term $\bf G$, which directly degrades the channel estimation accuracy.

Let us consider Corollary 1, then the dominant basis ${\bf D}^*_r$ for the CE-BEM can be rewritten as
\begin{equation}
{\bf D}^*_r = {\bf I}^{\langle q^*_r- \frac{Q}{2}\rangle}_K.  \label{eq20}
\end{equation}
Similarly, we have
\begin{align}
  {\bf H}_r 	& =  {\bf I}^{\langle q^*_r- \frac{Q}{2}\rangle}_K{\bf \Delta}^*_{r}, \label{eq29}
\end{align}
where ${\bf H}_r$ becomes a diagonal matrix with a permutation and its non-zero entries are corresponding to the dominant coefficients. From (\ref{eq29}), we find that, with Corollary 1, the desired signal is free of ICI but with a permutation of the received subcarrier. This is reasonable because the dominant coefficients in ${\bf c}^*_r$, corresponding to $f_r$, describe the channel alone while the non-dominant ones can be ignored. Therefore, by utilizing the position information, we can get the ICI-free pilots at the receive side and also reduce the needed channel coefficients from $KL$ to $L$.

\emph{Remark 1:} With Corollary 1, the conclusion that ${\bf H}_r$ is a permutated diagonal matrix only holds for the CE-BEM, since  ${\bf D}_q$ itself is a permutated identity matrix for the CE-BEM. For other BEMs, e.g., the GCE-BEM \cite{41}, the P-BEM \cite{42}, and the DPS-BEM \cite{43},  ${\bf D}_q$ is approximately banded. However, it can be expected that the proposed method can also highly reduce the ICI for other BEMs for only considering the dominant coefficients.

Assume ${\bf w}$ is received at the $r$-th receive antenna with the pilot pattern  ${\bf v}_r = [v_{r,1},v_{r,2},...,v_{r,P}]$.
Then, with Corollary 1, (\ref{eq26}) can be rewritten as
\begin{align}
{\bf y}_r({\bf v}_r)& =   {\bf D}^*_r({\bf v}_r,{\bf w}){\bf \Delta}^*_{r}({\bf w},{\bf w}){\bf x}({\bf w}) + \underbrace{ {\bf D}^*_r({\bf v}_r,{\bf d}){\bf \Delta}^*_{r}({\bf d},{\bf d}){\bf x}({\bf d})}_{{\bf G}^*} + {\bf n}_r({\bf v}_r),\\
&={\bf D}^*_r({\bf v}_r,{\bf w}){\bf \Delta}^*_{r}({\bf w},{\bf w}){\bf x}({\bf w}) +{\bf n}_r({\bf v}_r),\label{eq27}
\end{align}
where $  {\bf D}^*_r({\bf v}_r,{\bf w})$  and  ${\bf \Delta}^*_{r}({\bf w},{\bf w})$ represent the submatrices  with the row indices ${\bf v}_r$ and ${\bf p}$ and the column indices ${\bf p}$, respectively,
and ${\bf D}^*_{r}({\bf v}_r,{\bf d})$ and  ${\bf \Delta}^*_{r}({\bf d},{\bf d})$ represent the submatrices with row indices ${\bf v}_r$ and ${\bf d}$ and column indices ${\bf d}$, respectively. In (\ref{eq27}),
the term ${\bf G}^*$  denotes the  ICI caused by the data, and we have ${\bf G}^* = {\bf 0}$ since its corresponding entries of the dominant basis are zero, i.e., ${\bf D}^*_{r}({\bf v}_r,{\bf d}) = {\bf 0}$. 
Thus, it is easy to find the received pilots are free of the ICI but with a permutation of the received subcarriers.
The relationship between  ${\bf v}_r$ and ${\bf w}$ is given as
 \begin{equation}
  v_{r,p} = \left|w_p + (q^*_r-\frac{Q}{2})\right|_K, ~ w_p\in{\bf w}, ~  v_{r,p}\in{{\bf v}_r}, \label{eq23}
\end{equation}
where $p = 1,2,...,P$, and $|\cdot|_K$ denotes the mod $K$ operator.


\begin{figure}[t!]
  \centering
  \includegraphics[width=3.5in]{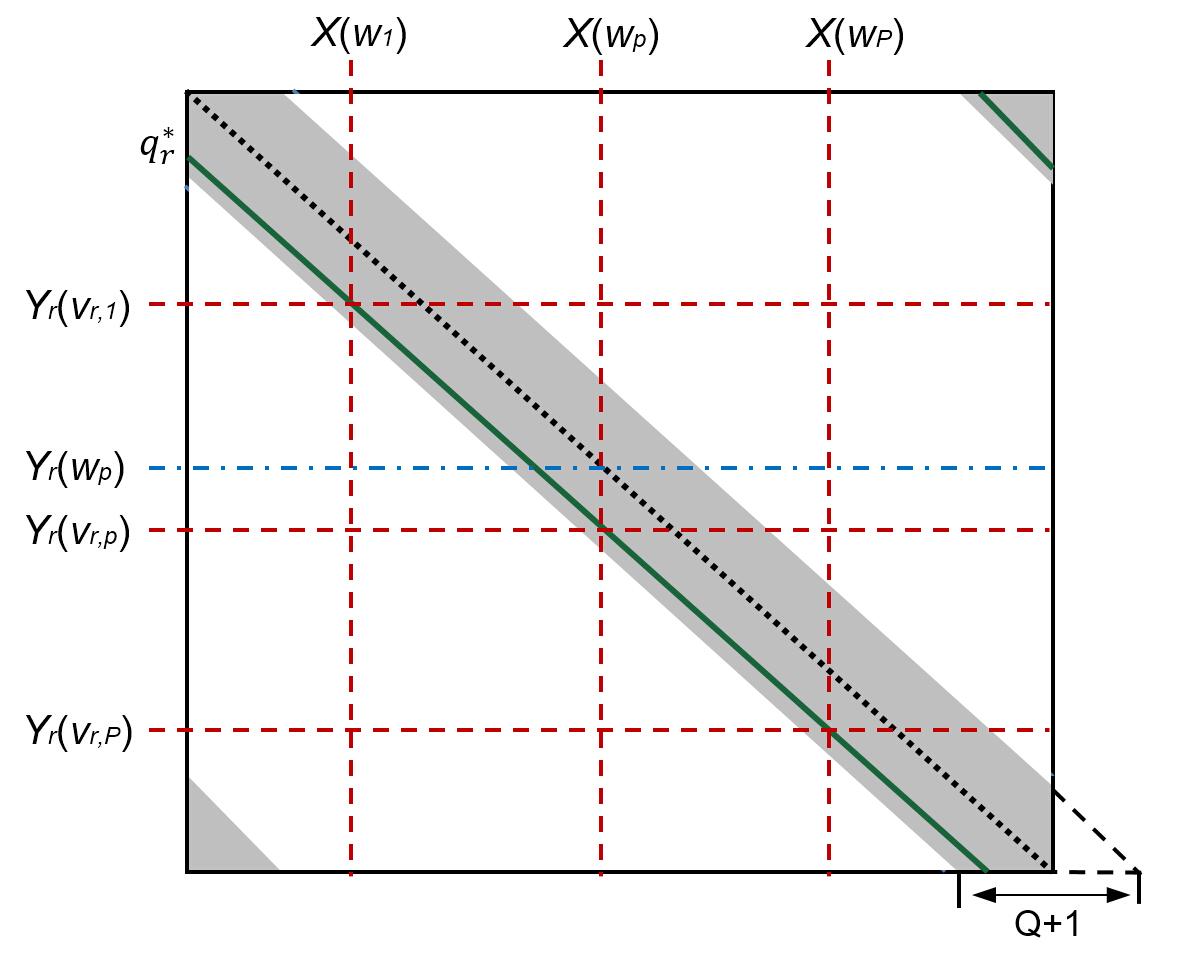}\\
  \caption{The structure of ${\bf H}_r$. (The grey parts denote the non-zero entries of ${\bf H}_r$, and the white parts denote the zero entries. The green solid line denotes the entries corresponding to the  dominant basis function ${\bf D}^*_r$ with the dominant index $q^*_r$. The black dot line denotes the diagonal entries of ${\bf H}_r$.)}\label{ici}
\end{figure}

For better clarification, we plot the structure of ${\bf H}_r$ in \reffig{ici}.  The columns of ${\bf H}_r$ are related to the subcarriers of the transmitted pilots and data, which operate on ${\bf D}_{q}$ through  ${\bf \Delta}_{r,q}$. The rows of ${\bf H}_{r}$ are related to the subcarriers of the received signals at the $r$-th receive antenna.
For the CE-BEM, ${\bf H}_r$ is strictly banded with the bandwidth $Q+1$, which is shown as the grey parts. From \reffig{ici}, it can be observed that a  received signal $Y_r(w_p)$ suffers from the ICI from the $Q$ neighbouring subcarriers of its desired signal $X(w_p)$, which is shown as the blue dash dot line.
Then, with Corollary 1, ${\bf H}_r$ turns to the green solid line and the grey parts can be neglected, which is because the dominant coefficients alone describe the channel with the Doppler shift $f_r$.
It is easy to find that the desired signal $X(w_p)$ is free of the ICI but received at $Y(v_{r,p})$  with a permutation of the received subcarrier, which is shown as the red dash lines.
Therefore, with the proposed method, the ICI among the received pilots at each receive antenna is eliminated.

\section{Low Coherence Compressed Channel Estimation}
In this section, based on  the proposed ICI elimination method, we design the pilot pattern to minimize the system average coherence and hence can improve the CS-based channel estimation performance.
First, we briefly review some fundamentals of CS. Then,  we formulate the problem and propose a new pilot pattern design algorithm to solve it.
Finally, we discuss the complexity and practical  applicability of our scheme.

\subsection{CS Fundamentals}
CS is an innovative technique to reconstruct sparse signals accurately from a limited number of measurements.
Considering  an unknown signal ${\hat{\bf{x}}}\in \mathbb{C}^M$, suppose that we have ${\hat{\bf{x}}}=\bf{\Phi}\bf{a}$, where ${\bf{\Phi}}\in {\mathbb{C}}^{M \times U}$ denotes a known dictionary matrix  and ${\bf a} \in {\mathbb{C}}^{U}$ denotes  a $S$-sparse vector, i.e., $\|{\bf{a}}\|_{{{{\ell }_{0}}}}=S\ll U$.
Then, CS considers the following problem
\begin{equation}
{\hat {\bf{{y}}}}={\bf{\Psi}}{\hat{ \bf{{x}}}} + {\bm{\eta}} =\bf{\Psi}\bf{\Phi}\bf{a} + {\bm{\eta}},
\end{equation}
in which ${\bf{\Psi}}\in {\mathbb{C}}^{V \times M}$ presents a known measurement matrix, ${\hat{\bf y}} \in {\mathbb{C}}^{V}$ presents the observed vector, and ${\bm{\eta}} \in  {\mathbb{C}}^{V}$ is the noise vector.
The objective of CS is to reconstruct  $\bf{a}$ accurately based on the knowledge of ${\hat{\bf y}}$, $\bf \Psi$, and  $\bf{\Phi}$.  It has been proved in  \cite{16}  that if $\bf {\Psi\Phi}$ satisfies the restricted isometry property (RIP) \cite{29}, then $\bf a$ can be reconstructed accurately with CS reconstruction methods, such as the basis pursuit (BP) \cite{30} and the orthogonal matching pursuit (OMP) \cite{31}.
In addition, a fundamental research \cite{17}  indicates that the average coherence reflects the actual CS behavior rather than the mutual coherence \cite{15} for considering the average performance. The definition of the average coherence is given as follows.

\begin{definition}[Average coherence  \cite{17}]
For a matrix $\bf{M}$ with the $i$-th column as ${\bf{g}}_i$, its average coherence is defined as the average of all absolute inner products between any two normalized columns in $\bf{M}$ that are beyond a threshold $\delta$, where $0<\delta<1$. Put formally
\begin{equation}
\mu _\delta \{ {\bf{M}}\}  = \frac{{\sum\limits_{i \ne j} {\left( {\left| {g_{ij} } \right| \ge \delta} \right) \cdot \left| {g_{ij} } \right|} }}{{\sum\limits_{i \ne j} {\left( {\left| {g_{ij} } \right| \ge \delta} \right)} }},
\end{equation}
where $g_{ij} = \tilde{\bf{ g}}^H_i \tilde{\bf{ g}}_j$, $\tilde {\bf{ g}}_i ={\bf{ g}}_i/\|{\bf{ g}}_i\|_{\ell_2} $, and
the operator is defined as
\begin{equation}
(x\geq y)=\left\{\begin{matrix}
  &1,~&x\geq y ,\\
  &0,~&x<y.
\end{matrix}\right.
\end{equation}
\end{definition}
 It has been established in \cite{17} that  a smaller $\mu_\delta \{\bf{\Psi\Phi}\}$ will lead to a more accurate recovery of $\bf a$. From this point of view, it can be expected that if $\bf{\Psi}$ is designed with a fixed $\bf{\Phi}$ such that ${{\mu_\delta}\left\{ {\bf{\Psi\Phi}} \right\}}$ is as small as possible, then CS can get better recovery performance.

\subsection{Problem Formulation}
To utilize the sparsity of the high-mobility channel according to Theorem 1, we  rewrite the received pilots at the $r$-th antenna as a function of channel coefficients.
In this paper, we assume that each receive antenna estimates its channel individually, and then sends the estimated channel to the RS for operation.
Then, (\ref{eq27}) can be rewritten as
\begin{align}
{\bf y}_r({{\bf v}_r})  ={\bf D}^*_r({\bf v}_r,{\bf w}){\bf S}{({\bf w},:)}{\bf c}^*_{r} +{\bf n}_r({\bf v}_r),
\end{align}
where ${\bf S}({\bf w},:) = {\text{diag}}\{{\bf x}({\bf w})\}{\bf F}_L({\bf w},:)$.
In this way, the task of estimating the high-mobility channel ${\bf H}_r$ in the frequency domain is converted to estimating the sparse coefficient vector ${\bf c}_r^*$.

As aforementioned in the pervious subsection, we have known that a lower $\mu_\delta$ leads to a better CS performance.
Therefore, we propose to design the pilot pattern ${\bf w}$ to minimize the average coherence in our system.
In this paper, we only design the pilot pattern and assume the pilot symbols are the same.
Therefore, the global pilot pattern design problem can be formulated as
\begin{equation}
 {\bf w}^* =  \arg \min_{\bf w}\max_{r} \mu_\delta\{{\bf D}_r^*({\bf v}_r,{\bf w}){\bf S}({\bf w},:)\}, \label{eq40}
\end{equation}
where ${\bf w}^*$ denotes the optimal  pilot pattern, and $r = 1,2,...,R$. Note that for a given ${\bf w}$, its corresponding ${\bf v}_r$ at the $r$-th receive antenna can be obtained by (\ref{eq23}). Thus, ${\bf w}$ is the only variable in this problem.

Taking the expression of ${\bf D}^*_r$ into consideration, the objective function can be represented as
\begin{align}\small
\mu_\delta\{{\bf D}_r^*({\bf v}_r,{\bf w}){\bf S}({\bf w},:)\}
&=  \mu_\delta\left\{ {\bf I}^{\langle {q^*_r}- \frac{Q}{2}\rangle}_K({\bf v}_r,{\bf w}){{\text{diag}}}\{ {\bf{x}}({{\bf{w}}})\}{{\bf{F}}_L}({{\bf{w}}},:)\right\}, \label{eq410}\\
&=  \mu_\delta\left\{ {{\text{diag}}}\{ {\bf{x}}({{\bf{w}}})\}{{\bf{F}}_L}({{\bf{w}}},:)\right\}, \label{eq41}
\end{align}
where we have ${\bf I}^{\langle {q^*_r}- \frac{Q}{2}\rangle}_K({\bf v}_r,{\bf w}) = {\bf I}_P$ for $r = 1,2,...,R$, and ${\bf I}_P$ denotes a $P\times P$ identity matrix. This is because  ${\bf v}_r$ and ${\bf w}$ are designed by the given equation (\ref{eq23}) to select the non-zero entries of ${\bf D}^*_r$.

Suppose that each pilot symbol has the same constant amplitude, i.e.,
\begin{equation}
|X({w}_{p})|^2 = A, ~~\forall{w}_{p}\in {\bf w}.
\end{equation}
According to Definition 2, it is not difficult to prove that  the average coherence is independent of the constant amplitude.
Thus, the objective function can be further written as
\begin{align}\small
\mu_\delta\{{\bf D}_r^*({\bf v}_r,{\bf w}){\bf S}({\bf w},:)\}
&=  \mu_\delta\left\{A {{\bf{F}}_L}({{\bf{w}}},:)\right\},\\
&=  \mu_\delta\left\{ {{\bf{F}}_L}({{\bf{w}}},:)\right\}.\label{eq44}
\end{align}

In this way, the problem (\ref{eq40})  is simplified to the following optimization problem
\begin{align}
 {\bf w}^* =  \arg \min_{{{\bf{w}}}} \mu_\delta\left\{ {{{\bf{F}}_L}({{\bf{w}}},:)}\right\}.\label{eq42}
\end{align}
From (\ref{eq42}), we find that the optimal pilot pattern ${\bf w}^*$ is independent of the train speed $v$, the Doppler shift $f_r$, the antenna number $R$, or the antenna position $\alpha_r$. This means that ${\bf w}^*$ is global optimal, regardless of the receive antenna number, the antenna position, the Doppler shift, or the train speed. Thus, for the given system, we can pre-design ${\bf w}^*$ and then sends it to each receive antenna to estimate the channel during the whole system runs. Note that the problem (\ref{eq42}) is different from the problem in our previous work \cite{25}, where the optimal pilot was related to the Doppler shift according to the instant train position.

\subsection{Low Coherence Pilot Pattern Design}

The similar pilot design problems for SISO-OFDM systems have been studied in \cite{26} and our previous work \cite{25}.
However, these methods cannot be directly applied to (\ref{eq42}). The problem in \cite{26} includes the guard pilots and needs to follow some constraints to eliminate the ICI. In addition, the optimal pilot in \cite{25} is related to the instant train position.
In this subsection, following the  spirit of \cite{25}, we propose a low complexity suboptimal pilot pattern design algorithm to solve this problem. The details are presented in Algorithm 1.

\begin{algorithm}[!t]\small
 \caption{{\bf{:}} Low Coherence Pilot Pattern Design}
  \begin{algorithmic}[1]

\Require Initial pilot pattern ${\bf w}$.
\Ensure Optimal pilot pattern ${\bf w}^*  = \hat{\bf w}^{(MP)}$.
\raggedright
\State {\bf Initialization}: Set $Iter = M\times P$, set ${\bf \Gamma}={\bf 0}$ and $\Gamma[0,0]=1$, set $\kappa=0$ and $\iota = 0$.

\For{$ n = 0, 1,..., M-1 $}
\For{$ k = 0, 1,..., P-1 $}
\State $m= n \times P+k$;
\\
~~~~~~~ {{a) Generate new pilot pattern}}:
        \State generate $\tilde{\bf{w}}^{(m)}$ with operator ${\bf{w}}^{(m)}\Rightarrow\tilde{\bf{w}}^{(m)}$;
        \If {$\mu_\delta\{ {{\bf{F}}_L}({\tilde{\bf{w}}^{(m)}},:)\} < \mu_\delta\{{{\bf{F}}_L}({{\bf{w}}^{(m)}},:)\}$}
         \State ${\bf{w}}^{(m+1)} = {{\tilde{\bf{w}}}^{(m)}}$;  $\kappa = m+1$;
        \Else
         \State ${{{\bf{w}}}^{(m+1)}} = {{{\bf{w}}}^{(m)}}$;
        \EndIf
        \\
~~~~~~~ {{b) Update state occupation probability and pilot pattern}}:
        \State ${\bf{\Gamma}}[m+1] = {\bf{\Gamma}}[m] + \eta[m]({\bf{U}}[m+1] - {\bf{\Gamma}}[m])$, with $\eta[m]=\frac{1}{m+1}$;
        \If {${{\Gamma}}[m+1,\kappa] > {{\Gamma}}[m+1,\iota]$}
            \State $\hat{\bf{w}}^{(m+1)} = {\bf{w}}^{({m+1})}$; $\iota \Leftarrow \kappa$;
        \Else
            \State $\hat{\bf{w}}^{({m+1})} = \hat{\bf{w}}^{({m})}$;
        \EndIf
\EndFor ~(k)
\EndFor ~(n)
\end{algorithmic}
\end{algorithm}

In Algorithm 1, ${\bf{w}}^{(m)}$, $\tilde{\bf{w}}^{(m)}$, and $\hat{\bf{w}}^{(m)}$ are defined as different pilot pattern sets  at the $m$-th iteration. $M$ is the number of  pilot pattern sets, and $Iter = M \times P$ denotes the total iteration times.
The probability vector ${\bf{\Gamma}}[m] = [\Gamma[m,1],\Gamma[m,2],...,\Gamma[m,MP]]^T$ represents the state occupation probabilities with entries ${{\Gamma}}[m,\kappa] \in [0,1]$, and $\sum_{\kappa}{{\Gamma}}[m,\kappa] =1$.
${\bf{U}}[m] \in \mathbb{R}^{MP\times 1}$ is defined as a zero vector except for its $m$-th entry to be 1.
In  Step a), $\tilde{\bf{w}}^{(m)}$ is obtained with the operator ${\bf{w}}^{(m)}\Rightarrow\tilde{\bf{w}}^{(m)} $, which is defined as: at the $m$-th iteration, the $k$-th pilot subcarrier of ${\bf{w}}^{(m)}$ is replaced with a random subcarrier which is not included in ${\bf{w}}^{(m)}$.
 Then, we compare ${{\tilde{\bf{w}}}^{(m)}}$ with ${{{\bf{w}}}^{(m)}}$ and select the one with a smaller system coherence to move a step.
In Step b), ${\bf{\Gamma}}[m+1]$ is updated based on the previous ${\bf{\Gamma}}[m]$ with the decreasing step size $\eta[m] = 1/(m+1)$. The current optimal pattern is updated by selecting the pilot pattern with the largest occupation probability.
Finally,  the optimal pilot pattern is obtained as ${\bf w}^* = \hat{\bf w}^{(MP)}$.
According to \cite{25}, this process can quickly converge to the optimal solution.

\emph{Remark 2:}
In contrast to the work in \cite{26},  Algorithm 1 does not need any guard pilot to eliminate the ICI, which highly improves the spectral efficiency. This is because the received pilots are ICI-free at each receive antenna with the proposed ICI elimination method. In specific, the total needed pilot number in \cite{26} is $(2Q+1)P$ ($P$ effective pilots and $2QP$ guard pilots), while our method only needs $P$ pilots.

\subsection{Complexity Analysis}
Here we briefly discuss the complexity of our proposed scheme. The complexity is mainly determined by the number of the needed complex multiplications. The complexity of the proposed scheme mainly consists of two parts: the low coherence pilot pattern design (Algorithm 1) and the position-based ICI elimination.
\begin{itemize}
	 \item For Algorithm 1,
it requires $MP^2(L(L-1) + M)$ complex multiplications in total.
In a practical system, as the constant parameters $M$, $L$, and $P$ are much smaller than $K$, the complexity of Algorithm 1 is much lower than  ${\mathcal{O}}(K^2)$.
Furthermore, since the needed system parameters can be estimated in advance, Algorithm 1 is an off-line process
and thus its complexity can be omitted in practice.
  \item For our ICI elimination method, with known the optimal pilot pattern  pre-designed by Algorithm 1, the $r$-th antenna obtains its receive pilot pattern by (\ref{eq23}) at any given position. This process only needs a permutation of the subcarriers, where the needed $q^*_r$ can be directly calculated from the HST's current speed and position information supported by the GPS. Thus, the proposed ICI elimination method introduces very low complexity in practical systems.
\end{itemize}

In addition, we also compare the system complexity of the proposed scheme and the scheme in our previous work \cite{25} in SIMO systems.
Note that the scheme in \cite{25} cannot directly extend to the SIMO system.
For SIMO systems, since the length of the HST cannot be ignored comparing with the cell range in practice,  the receive antennas may suffer from different Doppler shifts and correspond to different optimal pilots.
To solve this problem, based on the scheme in \cite{25}, one may divide the total $P$ pilots into $R$ subsets, and each subset sends the corresponding optimal pilot for each receive antenna to minimize the system coherence.  In the presence of  large number of the receive antennas (i.e., large $R$), this method will introduce high system complexity for selecting different optimal pilots. In addition, since the effective pilot number for each receive antenna is $P/R$, a large $R$ will also highly reduce the spectral efficiency for needing more total pilots to get satisfactory estimation performance.
However, for the proposed method in this work, since ${\bf w}^*$ is independent of the receive antenna position and receive antenna number, with increasing $R$, each receive antenna can still have $P$ effective pilots.

\subsection{Practical Applicability}
 Now we briefly discuss the applicability of the proposed scheme in a practical HST system.
 The entire process of our proposed scheme is summarized as follows:
 \begin{enumerate}
 	 \item  For a given HST system, as the system parameters can be collected in advance, the optimal pilot pattern ${\bf w}^*$ is pre-designed by Algorithm 1 and then pre-stored at both the BS and the HST. Since ${\bf w}^*$ is independent of the Doppler shift or the train position, the BS transmits ${\bf w}^*$ to estimate the channels during the whole process.  In contrast, in our previous work \cite{25},  the BS was required to select different optimal pilot for each receive antenna from a pre-designed codebook according to the  instant train position, which introduces high system complexity.
   \item Then, the antennas on the HST receive the signals and get the ICI-free pilots with the proposed ICI elimination method, which is given as (\ref{eq23}). In addition, with the instant train position and speed information supported by the GPS, ${q^*_r}$ of each receive antenna can be easily calculated with the given equations (\ref{q_fd}) and (\ref{q_a}).
   \item After that, each receive antenna uses the ICI-free pilots to estimate the channel coefficients with the conventional CS estimators.
 \end{enumerate}

In this way, the proposed scheme can be well used in current HST systems without adding too much complexity.
Note that, as ${\bf w}^*$ is also independent of the train speed, the performance of  the proposed scheme is robust to the high mobility. This is interesting because that the channel estimation performances are always highly influenced by the high system mobility \cite{25}\cite{32}. In the following section, we will give some simulation results to demonstrate the effectiveness of the proposed algorithm.

\section{Simulation Results}

\begin{table}[!t]
\centering
\caption{\label{table1} HST COMMUNICATION SYSTEM PARAMETERS} \label{tb1}
\renewcommand\arraystretch{1.1}
\begin{tabular}{c c c} \toprule
{Parameters} & {Variables}&{Values} \\  \toprule
BS cover range  & $R_{BS}$     &    $ 1000$~m      \\
HST length & $L_{hst}$ & $200$~m\\
Max distance of BS to railway  & $D_{max}$    &     $1000$~m \\
Min distance of BS to railway  & $D_{min}$    &     $40$~m     \\
Carrier frequency & $f_c$ & $2.35$~GHz \\
Train speed & $v$    &   $500$~km/h       \\
\bottomrule
\end{tabular}
\end{table}

In this section, we present the performance of the proposed scheme by two typical compressed channel estimators,  BP \cite{30} and OMP \cite{31}.
The mean square error (MSE) at each individual receive antenna and the bit error rate (BER) at the RS are illustrated versus the the signal to noise ratio (SNR) at different  HST positions.
We assume that the $R=2$ receive antennas are equipped, one at the front and the other at the end of the HST, respectively, i.e., the distance between the two receive antennas is equal to the HST length.
The HST system parameters are given in Table \ref{tb1}.
We consider a 512-subcarrier OFDM system with 40 pilot subcarriers, and the carrier frequency is $f_c=2.35$GHz.
The bandwidth is set to be $5$MHz, the packet duration is $T = 1.2$ms, and the modulation is $4$-QAM.
We consider the CE-BEM channel model and each channel has $L = 64$ taps, and only $5$ taps are dominant ones with random positions.
The speed of the HST is 500km/h, which means that the maximum Doppler shift is $f_{{max}} = 1.087$KHz. 
As a benchmark, the iterative ICI mitigation method in \cite{32} is simulated to compare with our proposed position-based ICI elimination method.

\subsection{MSE Performance}

\begin{figure}[!t]
\centering
\includegraphics[width=4in]{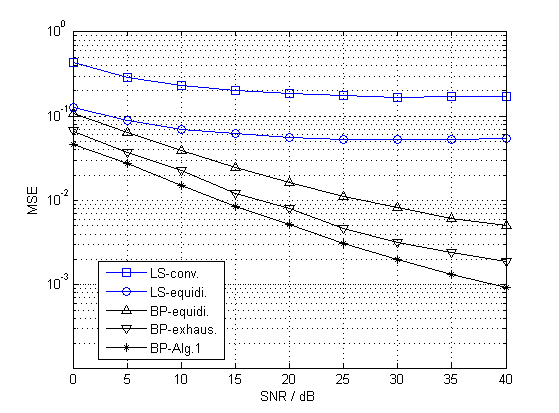}
\caption{MSE performances of the LS and the BP estimators with different pilot patterns at the position $A$.}\label{s1}
\end{figure}

\reffig{s1} gives the comparison of the MSE performances of different estimators with different pilot patterns at the position $A$, where the Doppler shift at the receive antenna is $1.087$KHz. In this figure, we compare three pilot pattern design methods.
The equidistant method (``equidi.'') is the equidistant pilot pattern in \cite{7}, which is claimed in \cite{7} as the optimal pilot pattern to doubly selective channels.
The exhaustive method (``exhaus.'') is the method in \cite{18} with $200$ iterations, which does an exhaustive search from a designed pilot pattern set.
The  iteration time of Algorithm 1 is set to be $Iter = 200$, which is shown in \cite{25} that $Iter = 200$ is good enough for a practical system.
The ``LS-equidi.'' method, the ``BP-equidi.'' method, the ``BP-exhaus.'' method, and the ``BP-Alg.1'' method are equipped with the proposed ICI elimination method.
In addition, the conventional LS method with the equidistant pilot pattern (``LS-conv.'') is equipped with the ICI mitigation method in \cite{32} with 2 iteration times.
It can be observed that  the BP  estimators significantly improve the performances than the LS methods by utilizing the sparsity of the high-mobility channels. Furthermore, it is found that the estimators with the proposed ICI elimination method get better performances than the one with the conventional method, which means that the proposed method effectively eliminates the ICI. As expected, comparing with other pilot patterns, Algorithm 1 improves the MSE performance for effectively reducing the system average coherence.

\begin{figure}[!t]
\centering
\includegraphics[width=4in]{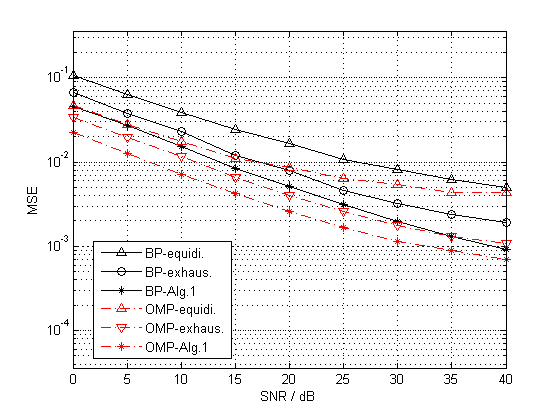}
\caption{MSE performances of the BP and the OMP estimators with different pilot patterns at the position $C$.}\label{s2}
\end{figure}

\reffig{s2} depicts the comparison of the MSE performances of BP and OMP estimators versus SNR with different pilot patterns at the position $C$, where the Doppler shift at the receive antenna is  $-1.087$KHz. All of BP and OMP estimators are considered with the proposed ICI elimination method. As can be seen, with Algorithm 1, both BP and OMP get better performances comparing with other pilot patterns. It can be seen that the proposed algorithm is effective to both BP and OMP estimators.

\begin{figure}[!t]
\centering
\includegraphics[width=4in]{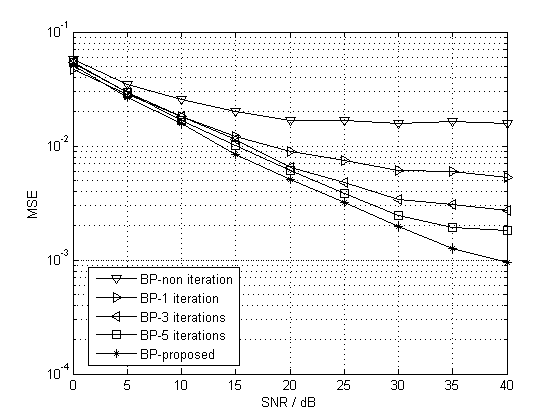}
\caption{MSE performances of the BP estimators with different ICI elimination methods.}\label{s3}
\end{figure}

\reffig{s3} presents the MSE performances of BP estimators versus SNR with the proposed ICI elimination method and the conventional ICI mitigation method,  where the Doppler shift at the receive antenna is $f_r = 1.009$KHz according to $\alpha_r = 900$m. The ICI mitigation is considered with the iteration time as 0, 1, 3, and 5 to show the performance tendency.  All of these estimators are equipped with the pilot pattern designed by Algorithm 1 ($Iter = 200$). It can be observed that, with increasing iterations, the BP with the ICI mitigation method converges to the one with the proposed ICI elimination method, which gets the ICI-free pilots as aforementioned analysis. In addition, we also notice that the ICI mitigation gain is limited with increasing iteration times due to the error propagation.

\begin{figure}[!t]
\centering
\includegraphics[width=4in]{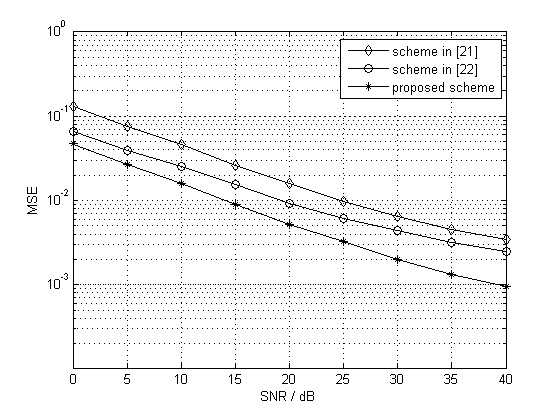}
\caption{Comparison of the MES performances of different schemes.}\label{s4}
\end{figure}

\reffig{s4} compares the MSE performances versus SNR of the proposed scheme, the scheme in \cite{25}, and the scheme in \cite{26},  where the Doppler shift at the receive antenna is $f_r =  1.087$KHz. The proposed scheme and the scheme in \cite{25} are both considered with 40 pilots and equipped with the proposed ICI elimination method.
However, since the optimal pilot in \cite{25} is related to the instant antenna position,  we divide the 40 pilots into two sets to send the optimal pilots for each antenna (20 effective pilots for each one).
In addition, the scheme in \cite{26} is considered with the guard pilots to get the ICI-free structure, and its total pilot number is $243$. Note that it needs $216$ guard pilots to eliminate the ICI, and thus its effective pilot number is $27$.
In this figure, it can be observed that the proposed scheme gets better performance with the same pilot number as \cite{25}. It is mainly because the optimal pilot for the proposed scheme is independent of the instant antenna position, i.e., the $R$ receive antennas correspond to the same optimal pilot, thus, each receive antenna has 40 effective pilots.
Furthermore, we notice that the proposed scheme is better than the scheme in \cite{26} with less pilot number. This is because the proposed ICI elimination method only needs a permutation of the receive subcarriers without needing any guard pilot, which highly improves the  spectral efficiency.

\begin{figure}[!t]
\centering
\includegraphics[width=4in]{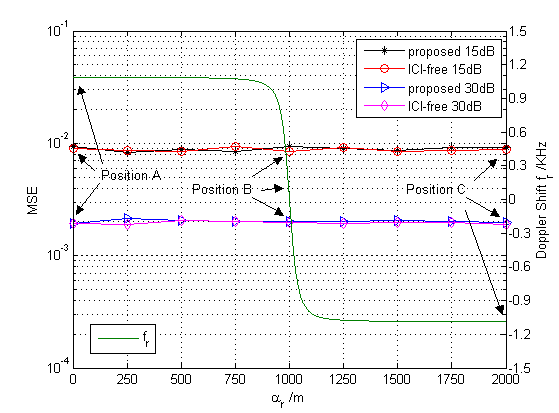}
\caption{MSE performances of BP estimators with Algorithm1 and the Doppler shift versus the antenna position $\alpha_r$.}\label{msevspos}
\end{figure}

\reffig{msevspos} presents the MSE performances of BP estimators versus the receive antenna position at SNR = $15$dB and SNR = $30$dB. As a reference, we also plot the Doppler shift $f_r$ at the $r$-th receive antenna versus its position $\alpha_r\in[0,2000]$m (from $A$ to $C$). We can find that $f_r$ changes from $f_{max}$ to $-f_{max}$ with the HST moves, and it changes rapidly near the position $B$.
In this figure, the resulting curves correspond to the performances when the proposed ICI elimination method is performed and when the estimation considers that pilots are free of ICI (``ICI-free'') at SNR= 15dB and 30dB, respectively. When the pilots are free of ICI, it means that the transmitted OFDM symbol is set as zero at the data subcarriers.
All estimators are considered with the pilot pattern designed by Algorithm 1 ($Iter = 200$).
From the curves, it can be observed that the proposed method and the ICI-free method are almost superimposed, which means that the proposed method can effectively obtain the ICI-free pilots.
In addition, we also notice that, although the HST suffers from large Doppler shift at most of the positions and $f_r$ changes rapidly near $B$, the MSE performances of the proposed method are stable.
This is because the optimal pilot pattern and the proposed ICI elimination method are both independent of the position information and the system mobility. Thus, the proposed scheme is robust with respect to high mobility.

\subsection{BER Performances}

\begin{figure}[!t]
\centering
\includegraphics[width=4in]{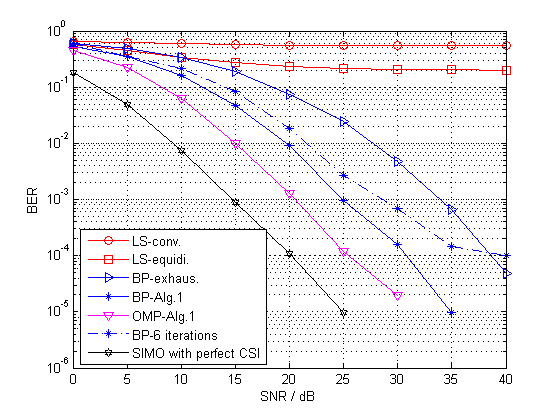}
\caption{BER performances of the $1\times2$ SIMO-OFDM system. }\label{ber1}
\end{figure}

\reffig{ber1} shows the BER performances versus SNR of the $1\times2$ SIMO-OFDM system in the given high-mobility environment at the position $A$, where the Doppler shifts at the receive antennas are both $1.087$KHz.
In this figure, we compare the LS, the BP, and the OMP estimators with the pilot patterns designed by the equidistant method, the exhaustive  method,  and Algorithm 1 ($Iter = 200$).
The conventional LS method (``LS-conv.'') is considered with the one-tap equalization, and other methods are considered with the zero-forcing (ZF) equalizer.
As a reference, we also plot the performance with the perfect knowledge of channel state information (CSI), which means that the CSI is available at the RS and employed with the ZF equalizer.
In addition, the BP estimator with Alg. 1 and the ICI mitigation method of 6 iteration times is also considered.
As can be seen, BP and OMP with Algorithm 1 are closer to the  perfect knowledge of CSI.
It is also shown that ``BP-Alg.1'' with the proposed ICI elimination method outperforms the one with the conventional ICI mitigation method for effectively eliminating the ICI. In addition, it can be observed that the pilot pattern designed by Algorithm 1 significantly improves the performances for effectively reducing the system coherence.

\begin{figure}[!t]
\centering
\includegraphics[width=4in]{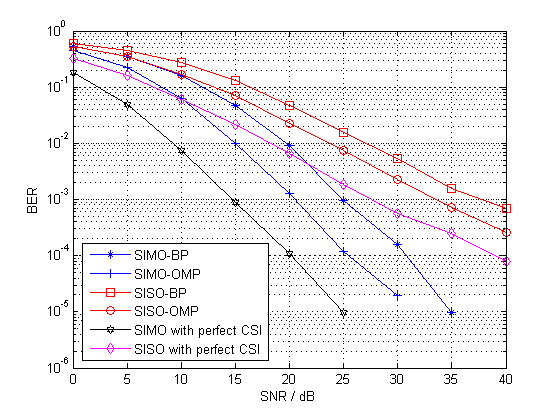}
\caption{BER performances of the SIMO and the SISO systems.}\label{ber2}
\end{figure}

\reffig{ber2} compares the BER performances between the $1\times2$ SIMO-OFDM system and the SISO-OFDM system in the given high-mobility environment at the position $A$.
Both of the SIMO and the SISO systems are considered with $40$ pilots. All estimators are equipped with the pilot pattern designed by Algorithm 1 and the proposed ICI elimination method.
As a reference, we plot the performances with the perfect knowledge of CSI for both the SIMO and  SISO systems.
It can be observed that the SIMO system significantly improves the BER performances due to the spatial diversity introduced by multiple antennas. 

\section{Conclusion}

In this paper, for the considered SIMO-OFDM HST communication system, we exploit the train position information and utilize it to mitigate the ICI caused by the high mobility. In specific, for the CE-BEM, we propose a new low complexity ICI elimination method to get the ICI-free pilots at each receive antenna. Furthermore, we design the pilot pattern to minimize the system coherence and hence can improve the CS-based channel estimation performance.
Simulation results demonstrate that the proposed scheme achieves better performances than the existing methods in the high-mobility environment.
In addition, it is also shown that the proposed scheme is robust to high mobility.




\begin{thebibliography}{40}



\bibitem{3}
Y. Mostofi and D. C. Cox, ``ICI mitigation for pilot-aided OFDM mobile systems," \emph{IEEE Trans. Wireless Commun.,} vol. 4, no. 2, pp. 765-774, March 2005.

\bibitem{4}
K. Kwak, S. Lee, H. Min, S. Choi, and D. Hong, ``New ofdm channel estimation with dual-ICI cancellation in highly mobile channel," \emph{IEEE Transactions on Wireless Communications}, vol. 9, no. 10, pp. 3155-3165, Oct. 2010.

\bibitem{5}
Z. Tang, R. C. Cannizzaro, G. Leus, and P. Banelli, ``Pilot-assisted time-varying channel estimation for OFDM systems," \emph{IEEE Transactions on Signal Processing,} vol. 55, no. 5, pp. 2226-2238, May 2007.

\bibitem{6}
H. Hijazi and L. Ros, ``Polynomial estimation of time-varying multipath gains with intercarrier interference mitigation in OFDM systems," \emph{IEEE Transactions on Vehicular Technology,} vol. 58, no. 1, pp. 140-151, Jan. 2009.

\bibitem{7}
X. Ma, G. Giannakis, and S. Ohno, ``Optimal training for block transmissions over doubly-selective wireless fading channels," \emph{IEEE Trans. Signal Process,} vol. 51, no. 5, pp. 1351-1366,  May 2003.


\bibitem{8}
W. U. Bajwa, A. M. Sayeed, and R. Nowak. ``Sparse multipath channels: modeling and estimation," \emph{IEEE Digital Signal Processing Education Workshop,} pp. 320-325, Jan. 2009.

\bibitem{9}
W. U. Bajwa, J. Haupt, A. M. Sayeed, and R. Nowak, ``Compressed channel sensing: a new approach to estimating sparse multipath channels," \emph{Proceedings of the IEEE}, vol. 98, no. 6, pp. 1058-1076, June 2010.

\bibitem{10}
W. U. Bajwa, A. M. Sayeed, and R. Nowak, ``Learning sparse doubly-selective channels," \emph{46th Annual Allerton Conference on Communication, Control and Computing,} pp. 575-582, Sept. 2008.


\bibitem{11}
G. Taubock and F. Hlawatsch, ``A compressed sensing technique for OFDM channel estimation in mobile environments: exploiting channel sparsity for reducing pilots," \emph{IEEE International Conference on Acoustics, Speech and Signal Processing (ICASSP),} pp. 2885-2888, March 2008.

\bibitem{12}
G. Taubock, F. Hlawatsch, D. Eiwen, and H. Rauhut, ``Compressive estimation of doubly selective channels in multicarrier systems: leakage effects and sparsity-enhancing processing," \emph{IEEE Journal of Selected Topics in Signal Processing,} vol. 4, no. 2, pp. 255-271, April 2010.


\bibitem{14}
C. Berger, Z. Wang, J. Huang, and S. Zhou, ``Application of compressive sensing to sparse channel estimation,'' \emph{IEEE Commun. Mag.}, vol. 48, no. 11, pp. 164-174, Nov. 2010.

\bibitem{15}
D. L. Donoho, M. Elad, and V. N. Temlyakov, ``Stable recovery of sparse overcomplete representations in the presence of noise," \emph{IEEE Trans. Inf. Theory,} vol. 52, no. 1, pp. 6-18, Jan. 2006.


\bibitem{16}
E. J. Candes, J. Romberg, and T. Tao, ``Robust uncertainty principles: exact signal reconstruction from highly incomplete frequency information,¡± \emph{IEEE Trans. Inf. Theory,} vol. 52, no. 2, pp. 489-509, Feb. 2006.

\bibitem{17}
M. Elad, ``Optimized projections for compressed sensing," \emph{IEEE Transcation on Signal Processing,} vol. 55, no. 12, pp. 5695-5702, Dec. 2007.


\bibitem{18}
X. He and R. Song, ``Pilot pattern optimization for compressed sensing based sparse channel estimation in OFDM systems," \emph{International Conference on Wireless Communications and Signal Processing (WCSP),} pp. 1-5,  Oct. 2010.

\bibitem{19}
N. Jing, W. Bi, and L. Wang, ``Deterministic pilot design for MIMO OFDM system based on compressed sensing," \emph{International Conference on Communication Technology (ICCT),} pp. 897-903, Nov. 2012.

\bibitem{20}
D. Wang and X. Hou, ``Compressed MIMO chanel estimation and efficient pilot pattern over Doppler sparse environment," \emph{International Conference on Wireless Communications and Signal Processing (WCSP),} pp. 1-5, Nov. 2011.

\bibitem{21}
C. Qi and L. Wu, ``Optimized pilot placement for sparse channel estimation in OFDM systems," \emph{IEEE Signal Processing Letters,} vol. 18, no. 12, pp. 749-752, Dec.  2011.

\bibitem{22}
C. Qi and L. Wu, ``A study of deterministic pilot allocation for sparse channel estimation in OFDM systems," \emph{IEEE Communications Letters,}  vol. 16, no. 5, pp. 742-744,  May 2012.

\bibitem{23}
X. Ren, W. Chen, and Z. Wang, ``Low coherence compressed channel estimation for high mobility MIMO OFDM  systems," \emph{Global Communications Conference (GLOBECOM),} Dec. 2013.


\bibitem{25}
X. Ren, W. Chen, and M. Tao, ``Position-based compressed channel estimation and pilot design for high-mobility OFDM systems," \emph{IEEE Transactions on Vehicular Technology,} July 2014. (Accepted)

\bibitem{26}
P. Cheng, Z. Chen, Y. Rui, Y.  Guo, L. Gui, M. Tao, and Q. T. Zhang ``Channel estimation for OFDM systems over doubly selective channels: a distributed compressive sensing based approach," \emph{IEEE Transaction on Communications,} vol. 61, no. 10,  pp. 4173-4185, Oct. 2013.

\bibitem{32}
H. Hijazi and L. Ros, ``Joint data QR-detection and kalman estimation for OFDM time-varying rayleigh channel complex gains," \emph{IEEE Trans. Commun.,} vol. 58, no. 1, pp. 170-178, 2010.


\bibitem{27}
L. Liu, C. Tao, J. Qiu, H. Chen, L. Yu, W. Dong, and Y. Yuan, ``Position-based modeling for wireless channel on high-speed railway under a viaduct at 2.35 GHz," \emph{IEEE Journal on Selected Areas in Communications,} vol. 30, no. 4, pp. 834-845, May 2012.

\bibitem{28}
R. D. Pascoe and T. N. Eichorn, ``What is communication-based train control?," \emph{IEEE Vehicular Technology Magazine,} vol. 4, no. 4, pp. 16-21, Dec. 2009.


\bibitem{251}
A. P. Kannu and P. Schniter, ``MSE-optimal training for linear time-varying channels,'' \emph{Proc. IEEE Int. Conf. Acoust., Speech, Singal Process. (ICASSP)}, pp. 789-792, Mar. 2005.




\bibitem{41}
G. Leus, ``On the estimation of rapidly time-varying channels,'' \emph{ Euro. Signal Process. Conf. (EUSIPCO)}, Sep. 2004.

\bibitem{42}
D. K. Borah and B. D. Hart, ``Frequency-selective fading channel estimation with a polynomial time-varying channel model,'' \emph{IEEE Trans. Commun.}, vol. 47, no. 6, pp. 862-873, Jun. 1999.

\bibitem{43}
T. Zemen and C. F. Mecklenbr\"{a}uker, ``Time-variant channel estimation using discrete prolate spheroidal sequences,'' \emph{IEEE Trans. Signal Process.}, vol. 53, no. 9, pp. 3597-3607, Sep. 2005.



\bibitem{29}
E. J. Candes and M. B. Wakin, ``An introduction to compressive sampling," \emph{IEEE Signal Processing Mag.,} vol. 25, no. 2,  pp. 21-30, March 2008.


\bibitem{30}
S. S. Chen, D. L. Donoho, and M. A. Saunders, ``Atomic decompostiion by basis pursuit," \emph{SIAM Review}, vol. 43, pp. 129-159, 2001.

\bibitem{31}
Y. C. Pati, R. Rezaiifar, and P. S. Krishnaprasad, ``Orthogonal matching pursuit: recursive function approximation with applications to wavelet decomposition," \emph{Proceedings of the 27th Annual Asilomar Conference on Signals, Systems and Computers}, vol. 1, pp. 40-44, Nov. 1993.







\end{thebibliography}
\end{document}